\documentstyle[epsf]{mn}
\title[Stellar capture by an accretion disc]{Stellar capture by an
 accretion disc}
\author[D.~Vokrouhlick\'y and
 V.~Karas]{D.~Vokrouhlick\'y$^{\,1,2\,}$\thanks{Since November 1997:
 Astronomical Institute, Charles University Prague,
 V~Hole\v{s}ovi\v{c}k\'ach~2, CZ-180\,00~Praha, Czech Republic.
 E-mail: vokrouhl@mbox.cesnet.cz; karas@mbox.cesnet.cz} and
 V. Karas$^{\,1,3\,\mbox{$\star$}}$\\
 $^{1}$Astronomical Institute, Charles University Prague, \v{S}v\'edsk\'a 8,
  CZ-150\,00~Praha, Czech Republic\\
 $^2$Observatoire de la C\^ote d'Azur, dept.\ CERGA, Av.\ N.\ Copernic,
  F-06130~Grasse, France\\
 $^3$Scuola Internazionale Superiore di Studi Avanzati, Via Beirut 2/4,
  I-34014~Trieste, Italy}
\date{Received \today}
\pagerange{\pageref{firstpage}--\pageref{lastpage}}
\newcommand{\etal}{{\rm et~al.\ }}
\newcommand{\myder}{{\mbox{d}}}

\font\tenmsbm=msbm10
\font\sevenmsbm=msbm7
\font\fivemsbm=msbm5
\newfam\msbmfam
\textfont\msbmfam=\tenmsbm
\scriptfont\msbmfam=\sevenmsbm
\scriptscriptfont\msbmfam=\fivemsbm
\def\akpa{\mathchar"2B7B} 

\begin{document}
\label{firstpage}
\maketitle

\begin{abstract}
 Long-term evolution of a stellar orbit captured by a massive galactic
 center via successive interactions with an accretion disc has been
 examined. An analytical solution describing evolution of the stellar
 orbital parameters during the initial stage of the capture was found.
 Our results are applicable to thin Keplerian discs with an arbitrary
 radial distribution of density and rather general prescription for the
 star-disc interaction. Temporal evolution is given in the form
 of quadrature which can be carried out numerically.
\end{abstract}


\section{Introduction}
Evolution of the orbit of a star under the influence of interactions
with an accretion disc has been studied by numerous authors because this
situation is relevant to inner regions of active galactic nuclei.
The trajectory of an individual star is determined mainly by gravity of
the central mass and surrounding stars while periodic transitions
through the disc act as a tiny perturbation. The final goal is to
understand the fate of a star, transfer of mass and angular momentum
between the star and the disc, and also to determine how star-disc
interactions influence the distribution of stellar orbits near a massive
central object. An important and difficult task is to estimate the
probability that a star gets captured from an originally unbound orbit,
and to determine whether this probability is significant compared to
other mechanisms of capture.

Orbital evolution of a body crossing an accretion disc has been
discussed with various approaches, first within the framework of
Newtonian gravity, both in theory of the solar system (Pollack, Burns \&
Tauber 1979; McKinnon \& Leith 1995) and for active galactic
nuclei (Goldreich \& Tremaine
1980; Syer, Clarke \& Rees 1991; Artymowicz 1994; Podsiadlowski \& Rees
1994). These studies have been generalized in order to account for the
effects of general relativity \cite{VK93} and to model a dense star
cluster in a galactic nucleus \cite{PL94,R95}. It has been recognized
that detailed physical description of the star-disc interaction is a
difficult task (Zurek, Siemiginowska \& Colgate 1994, 1996). In this
letter a simplified analytical treatment of stellar orbital parameters
is presented, describing the initial stage of star-disc collisions (when
the star crosses the disc once per revolution). A great deal of
our calculation is independent of microphysics of star-disc
interaction. It is shown how our
solution matches the corresponding Rauch's \shortcite{R95} solution
which is valid in later stages, when eccentricity of the orbit becomes
small enough.

In the next section our approach to the problem is formulated and an
analytical solution is given. Then, in Sec.~\ref{details}, further
details about the derivation of the results are presented, and finally a
simple example of the orbital evolution is shown in Sec.~\ref{example}.

\section{Stellar capture by a disc}
\subsection{Formulation and results}
\label{formulation}
Newtonian gravitational law is assumed throughout this paper. Our solution
is based on the following assumptions:

\begin{description}
\item[(i)] the disc is geometrically thin and its rotation is Keplerian;
\item[(ii)] at the event of crossing the plane of the disc, velocity of the
 star is changed by a tiny quantity. This impulse is colinear with the
 relative velocity of the star with respect to the material forming the disc;
\item[(iii)] the star crosses the disc once per revolution
 (the model is applicable to the initial phase of the stellar capture).
\end{description}
The first item is a standard simplification in which the disc is treated
in terms of vertically integrated quantities, while the second one can be
expressed by the formula for an impulsive change of the star's velocity:

\begin{equation}
 \delta\bmath{v}=\Sigma(\bmath{r},\bmath{v})\,\bmath{v}_{\rm rel}\,;
 \label{2one}
\end{equation}
$\Sigma$ is an unconstrained function, given by a detailed model of the
star-disc interaction, and $\bmath{v}_{\rm rel}$ is relative velocity of
the star and the disc material. We stress, that our results are uniquely
based on this assumption of colinearity,
$\delta\bmath{v}\propto\bmath{v}_{\rm rel}$; the coupling factor
$\Sigma$ is arbitrary and it can be as complex as necessary. In
particular, $\Sigma$ contains information about the surface density
$\akpa$ of the
disc ($\akpa=0$ outside an outer edge $r=R_{\rm{d}}$ of the disc). The
form of $\Sigma$ must be specified only for examination of temporal evolution
of orbital parameters. We will assume, in analogy with Rauch
(1995), a simplified formula for

\begin{eqnarray}
 \Sigma(\bmath{r},\bmath{v}) &=& -{\pi R_\star^2\over m_\star}\,
 \akpa(r)\,{v_{\rm rel}\over v_\perp}\,\xi\,,
 \label{22two} \\
 \xi &\approx& 1+\left({v_\star\over v}\right)^4\ln\Lambda
 \,,\label{sigma}
\end{eqnarray}
when it is needed for purpose of an example. In eq.~(\ref{22two})
$R_\star$ denotes radius of the star, $m_\star$ is its mass,
$v_\star$ escape velocity ($v^2_\star=2Gm_\star/R_\star$);
$v_\perp$ is normal component of the star's velocity to the disc plane,
and $\ln\Lambda$ is a usual long-range interaction factor (Coulomb
logarithm).

The star's orbit is traditionally characterized by the Keplerian
osculating elements: semimajor axis $a$, eccentricity $e$, inclination
$I$ to the accretion disc plane, and longitude of pericenter $\omega$
(measured from the ascending node). A derived set of parameters turns
out to be better suited to our problem: $\alpha=1/a$,
$\eta=\sqrt{1-e^2}$, $\mu=\cos{I}$, and $k=e\cos\omega$. We will show
(Sec.~\ref{details}) that evolution of a stellar orbit following
the capture by a disc can be written in a parametrical form:

\begin{eqnarray}
 \alpha(\zeta) &=& \phi(\zeta)\left\{\alpha_0 \phi^{-1}(\zeta_0) +
  \sigma^2\left[\psi(\zeta)-\psi(\zeta_0)\right]\right\}\,,
 \label{one} \\
 \eta^2(\zeta) &=& \zeta\phi(\zeta)\left\{\alpha_0 \phi^{-1}(\zeta_0)
  \sigma^{-2} + \left[\psi(\zeta)-\psi(\zeta_0)\right]\right\}\,,
 \label{two}\\
 \mu(\zeta) &=& \sqrt{\zeta} + \theta(\zeta)\,, \label{three}\\
 \mid\!k(\zeta)\!\mid{} &=& \zeta -1\,, \label{four}
\end{eqnarray}
where the auxiliary functions $\phi(\zeta)$, $\theta(\zeta)$ and
$\psi(\zeta)$ read
\begin{eqnarray}
 \phi(\zeta) &\!=\!& {1\over C} \left(1\pm \sqrt{1-C+C\zeta}
  \right)^2\,, \label{five}\\
 \theta(\zeta) &\!=\!& \mp{1\over C} \sqrt{{1-C+C\zeta\over \zeta}}
  \left(1\pm \sqrt{1-C+C\zeta}\right)\,, \label{six}\\
 \psi(\zeta) &\!=\!& {1\over 1\pm\sqrt{1-C+C\zeta}}\left(2+{C\over 1\pm
  \sqrt{1-C+C\zeta}}\right). \label{seven}
\end{eqnarray}
Formal parameter $\zeta$ of the solution decreases from its
initial value $\zeta_0=1+e_0\mid\!\cos\omega_0\!\mid$ to the final
value $\zeta_{\rm f}$, given by
\begin{equation}
 \zeta_{\rm f} = {2 R_{\rm{d}}\sigma^2\over 1+R_{\rm{d}}\sigma^2}
  \,. \label{fourteen}
\end{equation}
At this instant, the orbit starts crossing the disc twice per revolution
and our solution ceases to be applicable. Obviously, integration
constants in (\ref{one})--(\ref{seven}) are determined in terms of the
initial Keplerian orbital elements $(a_0,e_0,I_0,\omega_0)$ by

\begin{eqnarray}
 \alpha_0 &=& {1\over a_0}\,, \label{eight}\\
 \zeta_0 &=& 1 + e_0\mid\!\cos\omega_0\!\mid{}\,, \label{nine}\\
 \sigma^{-2} &=& {a_0\eta_0^2\over \zeta_0}\,, \label{ten}\\
 C &=& {z_0 (z_0 +2)\over (z_0+1)^2} {1\over 1-\zeta_0}\,,
 \label{eleven}
\end{eqnarray}
where
\begin{equation}
 z_0 = -{1-\zeta_0\over \sqrt{\zeta_0}(\cos I_0 -
  \sqrt{\zeta_0})}\,.\label{twelve}
\end{equation}
Upper signs in (\ref{five})--(\ref{seven}) are for the initial
inclination $I_0$ greater than a critical value
$I_\star$ given by

\begin{equation}
 \cos I_\star={1\over \sqrt{\zeta_0}}\,,\label{thirteen}
\end{equation}
lower signs apply otherwise. Integration constant $C$ is singular ($C
\rightarrow \infty$) for $I_0=I_\star$ ($z_0=-1$), and the solution
can be simplified further. For instance,
$\mu(\zeta)=1/\sqrt{\zeta}$ for all values of $\zeta$
down to $\zeta_{\rm f}$.

Solution (\ref{one})--(\ref{four}) can be extended easily to the
case of initially parabolic orbits by setting $\alpha_0=0$,
$e_0=1$ and $\sigma^2=(\zeta_0/2 R_{\rm p})$. Here, $R_{\rm p}$
denotes pericenter distance of the initial parabolic orbit.

It is worth mentioning that the parameter $\zeta$ does not
determine the time-scale on which the evolution takes place. Indeed,
eqs.~(\ref{one})--(\ref{four}) do not provide temporal
information because it depends on the precise form of
$\Sigma$ in eq.~(\ref{2one}). On the other hand, the strength
and the beauty of the parametric solution (\ref{one})--(\ref{four}) is
in its independence on a particular model for $\Sigma$.
We will also illustrate an example of temporal evolution
later in the text, and
only for this purpose the form of $\Sigma$ will be needed.
Assuming relation (\ref{22two})--(\ref{sigma}) one obtains

\begin{equation}
t-t_0=\frac{\pi}{\sigma^3\sqrt{GM}\Sigma_{\rm{c}}}\int_\zeta^{\zeta_0}
 \frac{\sigma^3\myder{z}}{z^{1/2}\alpha^{3/2}(z)\nu(z)\theta(z)}
 \label{time}
\end{equation}
where $t_0$ is initial time, $M$ is the central mass, and
$\Sigma_{\rm{c}}=(\pi{}R_\star^2/m_\star)\akpa(r_{\rm{c}})\xi$
with $r_{\rm{c}}=\sigma^{-2}$ being radial distance of the point
of intersection with the disc. Function $\nu=v_{\rm{rel}}/v_\perp$
is determined by orbital parameters which themselves depend on
$\zeta$ according to eqs. (\ref{one})--(\ref{four}).

We note that Rauch \shortcite{R95} conjectured that function
$R=a\eta^2\cos^4(I/2)$ remains nearly conserved along the evolving stellar
orbit and he used
this function for estimates of the radius of the final, circularized
orbit in the disc plane. In our notation,

\begin{equation}
 R(\zeta)={\zeta\over 4\sigma^2}\left[1+\sqrt{\zeta}+\theta
  \left(\zeta\right)\right]^2.
  \label{fifteen}
\end{equation}
Hereinafter, we show that $R(\zeta)$ is a
well-conserved quantity at later stages of the orbital evolution (when
eccentricity is sufficiently small), but it fails to be conserved at the
very beginning of the capture when the orbit is still nearly parabolic,
close to an unbound trajectory.

\begin{figure}
 \epsfxsize=\hsize
 \centering
 \mbox{\epsfbox[14 40 581 470]{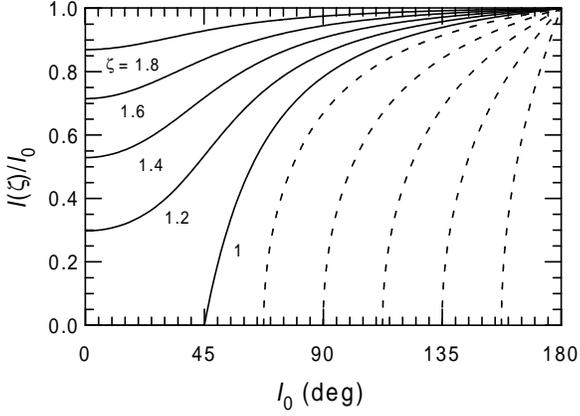}}
 \caption{Inclination $I(\zeta)$ of a captured stellar orbit (measured
  in units of initial inclination $I_0$) vs.\ $I_0$. This graph
  corresponds to initially parabolic orbits with pericenter distance
  equal to the disc radius. The curves are parametrized by $\zeta$.
  Temporal evolution of some particular orbit starts with $I=I_0,$
  $\zeta=2,$ and goes down along the vertical line, across $\zeta={\it
  const}$ curves. Our analytic solution (solid lines)
  is valid in the region $2\geq\zeta\geq1$, and it
  corresponds to eccentric orbits. Circularization time-scale is equal
  to the time-scale necessary for grinding the orbit to the disc plane
  if $I_0<I_\star=45\degr\!\!$, while the former is shorter than the
  latter for orbits with $I_0>I_\star$. At $\zeta=1$, i.e.\ zero
  eccentricity, our solution matches Rauch's $R={\it const}$
  solution (dashed lines).
 \label{fig1}}
\end{figure}

\subsection{Details of the solution}
\label{details}
In this section, we present more details of the derivation of the
solution given above.

Taking into account the fundamental assumption (\ref{2one}), one can
easily demonstrate that the Keplerian orbital elements are perturbed
at each transition (due to interaction with the disc material) by
quantities
\begin{eqnarray}
 \delta (\sqrt{a}) &=& \Sigma\, \sqrt{a}\eta\, f_1 \,,
  \label{2two}\\
 \delta (\sqrt{a}\eta) &=& \Sigma\, \sqrt{a}\eta\, f_2 \,,
  \label{2three}\\
 \delta (\sqrt{a}\eta\mu) &=& \Sigma\, \sqrt{a}\eta\, f_3 \,,
 \label{2four}\\
 \delta (k) &=& 2\Sigma\, (1+k)\, f_2\,. \label{2five}
\end{eqnarray}
Here, we introduced auxiliary functions
\begin{eqnarray}
 f_1 &=& \eta^{-3}\left\{\left(1+k\right)\left[1-\mu\sqrt{1+k}
  \right] + e^2 + k\right\}\,, \label{2six} \\
 f_2 &=& 1-{\mu\over \sqrt{1+k}}\,, \label{2seven}\\
 f_3 &=& \mu-{1\over \sqrt{1+k}}\,. \label{2eight}
\end{eqnarray}
Combining eqs.~(\ref{2three}) and (\ref{2five}) we find that
\begin{equation}
 {\sqrt{1+k}\over \sqrt{a}\eta} = \sigma \label{2nine}
\end{equation}
is conserved at the star-disc interaction. Hence, $\sigma$ is constant
whatever the evolution of elements $a$, $e$ and $k$ is. In fact,
condition (\ref{2nine}) states that the initial Keplerian orbit has the
same radius of intersection as the final orbit, after successive interactions
with the disc. Longitude of the node is also conserved and can be set to
zero without loss of generality. The above-given formulae
(\ref{2two})--(\ref{2nine}) correspond to $k>0$ (i.e.\
$\mid\!\omega\!\mid{}<\pi/2$); for $k<0$ one should replace
$k\rightarrow-k$. We note that all these relations can be easily
reparametrized in terms of binding energy $E=GM/(2a)$, angular momentum
$L=\sqrt{GMa}\eta$, and component of the angular momentum with respect
to axis, $L_{\rm{z}}=\sqrt{GMa}\eta\mu$.

Combining eqs.~(\ref{2three}) and (\ref{2four}) with the help of
(\ref{2nine}), and introducing auxiliary variables $y=\sqrt{a}\eta\mu$
and $x=\sqrt{a}\eta$, we obtain differential equation

\begin{equation}
 {\myder{y}\over\myder{x}}={x(\sigma{}y-1)\over\sigma{}x^2-y}\,.
 \label{2ten}
\end{equation}
(We were allowed to change variations, $\delta$, to the differentials,
$\myder$,
assuming an infinitesimally small perturbation of the stellar orbit at
each intersection with the disc.) The Abel-type differential equation
(\ref{2ten}) can be solved beautifully by standard methods of
mathematical analysis (see, e.g., Kamke 1959). An appropriate change of
variables gives directly a solution for the evolution of inclination,
eq.~(\ref{three}).

Similarly, considering (\ref{2two}) and (\ref{2three}) in terms
of $\alpha=1/a$, we obtain, after brief algebraic transformations,
\begin{equation}
 -\sqrt{\zeta}\theta(\zeta)\, {\myder\alpha\over\myder\zeta} +\alpha
  (\zeta) = \sigma^2 \left[2-\zeta+\sqrt{\zeta}\theta\left(\zeta
  \right)\right] \label{2eleven}
\end{equation}
with $\theta(\zeta)$ given by eq.~(\ref{six}). This is a linear
differential equation, integration of which yields $\alpha(\zeta)$ and
then, by eqs.~(\ref{one})--(\ref{two}), also $\eta(\zeta)$.

At this point, one can see that Rauch's \shortcite{R95} ``quasi-integral''
$R$ is changed at each passage across the disc
according to

\begin{equation}
 \delta(\ln R)=2\Sigma\left(1-{1\over\sqrt{1+k}}\right)\,.
 \label{2twelve}
\end{equation}
Realizing that $k\approx e$ we conclude that $\ln R$ indeed stays nearly
constant at later stages of the orbit evolution, when eccentricity has
decreased enough. On the other hand, at the very beginning of the
capture process, when eccentricity is still high, $R$ fails to serve as a
quasi-integral of the problem. Instead, its evolution is given by
eq.~(\ref{fifteen}).

For temporal evolution, eqs.~(\ref{2two})--(\ref{2nine}) must be
supplemented by additional relation,

\begin{equation}
\delta(t)=\frac{2\pi}{\sqrt{GM}}\,a^{3/2},
\label{deltat}
\end{equation}
which determines interval between successive intersections with
the disc. Combining eq.~(\ref{deltat}) with (\ref{2nine}) one
obtains separated differential equation which yields formula
(\ref{time}). Recall that this last step requires assumption
(\ref{22two}) about the form of $\Sigma$. In the present case,

\begin{equation}
\nu(\zeta)=\frac{1}{\zeta}
\sqrt{\frac{2\zeta(1-\zeta-\sqrt{\zeta}\theta)+\zeta-\eta^2}{1-
\zeta-2\sqrt{\zeta}\theta-\theta^2}}.
\end{equation}
Relation for time is apparently too complicated to be integrated
analytically but numerical evaluation is straightforward.

\begin{figure}
 \epsfxsize=\hsize
 \centering
 \mbox{\epsfbox[14 40 581 470]{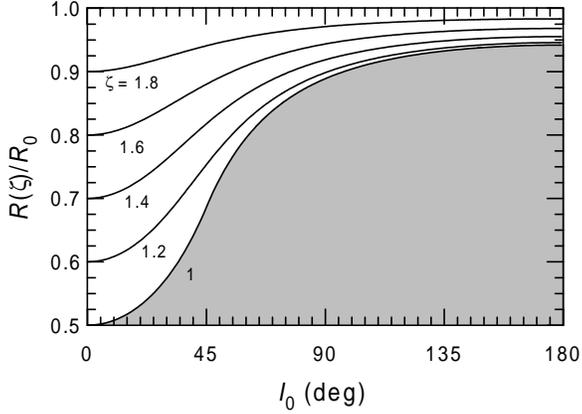}}
 \caption{Function $R$ (normalized to its initial value
  $R_0$) vs.\ initial inclination $I_0$ of the orbit for different
  values of parameter $\zeta$. The $\zeta$-dependence is stronger
  for orbits with small $I_0$. Once the orbit becomes circular with
  $\zeta=1$, $R$ reaches its terminal value and does not change
  any more. Therefore, $R$ does not acquire values inside
  the shaded region.
 \label{fig2}}
\end{figure}

\subsection{Example}
\label{example}
We shall briefly demonstrate some properties of the analytical solution
from Sec.~\ref{formulation}.

We examine parabolic orbits ($\alpha_0=0$, $e_0=1$) with pericenter in the
disc plane ($\omega_0=0$), and the pericenter distance $R_{\rm p}$ equal
to the disc radius ($R_{\rm p}=R_{\rm d}=r_{\rm{c}}$).
Initial inclination $I_0$ of
the orbit to the disc plane is a free parameter in this example.
Evolution of this set of orbits is split into two phases.

First, we let the orbits evolve according to the solution of
eqs.~(\ref{one})--(\ref{four}) from the initial value $\zeta_0=2$ of the
formal parameter $\zeta$ to its final value $\zeta_{\rm f}=1$.
Figure~\ref{fig1} illustrates the evolution of the inclination
$I(\zeta)$, measured in terms of the initial value $I_0$. Notice that
the critical inclination $I_\star$ of eq.~(\ref{thirteen}) is $45\fdg$
Eccentricity of the orbits under consideration decreases according to a
simple formula $e(\zeta)=\zeta-1$ (independently of $I_0$), leading
eventually to circularized orbits at $\zeta=\zeta_{\rm f}$. We observe
that orbits with $I_0<I_\star$ terminate at the final state $I_{\rm f}=0$,
suggesting that the circularization time-scale is comparable to
that necessary for grinding the orbit into the disc plane. On the other
hand, when $I_0>I_\star$ the final circular orbits remain inclined
significantly to the disc plane. ($I>90\degr$ corresponds to
retrograde orbits.) Hence, for those orbits the
circularization time-scale is shorter than the time necessary for
tilting the orbit to the disc plane. Additional time
to incline a circularized orbit is not much longer than circularization
time, however.\footnote{We thank the referee for pointing out this fact,
confirmed also by other estimates \cite{SCR91,KL95}.} The difference is
typically a factor of 10 for highly retrograde orbits.

Secondly, we examine the evolution of circularized orbits which started
with $I_0>I_\star$ and have settled to nonzero $I(\zeta=1)$. Because
these orbits have zero eccentricity, there exists Rauch's integral in
the form $R_1=\sigma^2{a}\cos^4(I/2)=\zeta\cos^4(I/2)$. Here, we adopted
a formal continuation of the $\zeta$ parameter to values smaller than
unity (in this phase, $\zeta=a/R_{\rm d}$). For each orbit, we calculate
the value of $R_1\equiv{}R(\zeta=1)$, so that the inclination is given
by

\begin{equation}
 \mu(\zeta)=\sqrt{{4R_1\over \zeta}}-1\,. \label{3one}
\end{equation}
Obviously, a given orbit terminates its evolution at $\zeta=4R_1$
when it is pushed completely into the disc plane.
Dashed curves in Figure~\ref{fig1} correspond to constant
values of $\zeta<1$.

Figure~\ref{fig2} illustrates how function $R(\zeta)$ changes during
the first circularization phase of the evolution. For each orbit we have
chosen the same steps in $\zeta$ (0.2) in the range
$1.8\geq\zeta\geq1.0$, as in Figure~\ref{fig1}, and we computed the
corresponding values of $R(\zeta)$ from eq.~(\ref{fifteen}). Our results
agree with Rauch's \shortcite{R95} finding, namely that $R(\zeta)$ is
conserved up to a factor of $\approx2$ for orbits with large eccentricity.
During the second phase of the evolution the $R$-function is constant.

\begin{figure}
 \epsfxsize=\hsize
 \centering
 \mbox{\epsfbox[14 40 581 470]{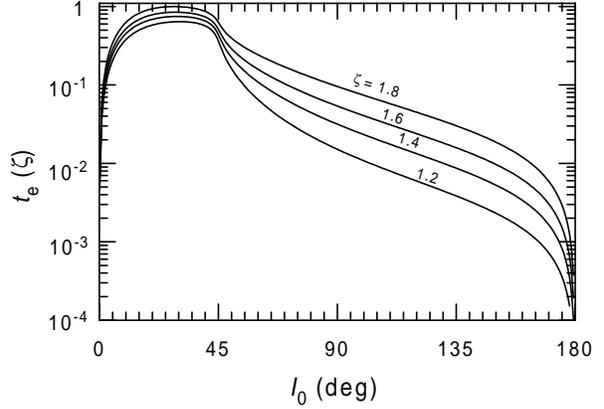}}
 \caption{Time $t_{\rm{e}}(\zeta)$ of orbital circularization of
parabolic orbits with initial inclination $I_0$, as in Fig.~1.
Here, time (arbitrary units on vertical axis) is recorded starting
from eccentricity $e=\zeta-1$ (given with each curve) down to
$e=10^{-3}$ (nearly circular orbit). Notice the
change in form of the curves at critical inclination $I_\star=45\fdg$
 \label{fig3}}
\end{figure}

Figure~\ref{fig3} shows time intervals $t_{\rm{e}}(\zeta)$ which elapse
in the course of gradual circularization when eccentricity
decreases from $e=\zeta-1$ to some terminal value (here, terminal
eccentricity has been fixed to $e=10^{-3}$; notice that $t_{\rm{e}}$ goes
to infinity for terminal eccentricity $e\rightarrow0$). We have
verified the graph also by direct numerical integration of the
corresponding orbits. Numerical factor standing in front of integral
on the right-hand side of eq.~(\ref{time}) can be written in physical
units in the form

\begin{equation}
10^7\left(\frac{r_{\rm{c}}}{10^3\,R_{\rm{g}}}\right)^{9/4}
\left(\frac{R_{\rm{g}}}{10^5R_\star}\right)
\left(\frac{10^3R_{\rm{g}\star}}{R_\star}\right)
\left(\frac{10^3}{\xi}\right)\,\mbox{yrs};
\end{equation}
$R_{\rm{g}}=2GM/c^2$ and $R_{\rm{g}\star}=2Gm_\star/c^2$ are gravitational
radii of the central mass and the star, respectively.
A typical surface density profile of the disc has been assumed, as in
eq.~(1) of Rauch (1995). The value of $\xi\approx10^3$
corresponds to the estimate in addendum to Zurek, Siemiginowska \& Colgate
(1996).

\section{Conclusion}
We have found a solution describing the evolution of orbital parameters
of a star orbiting around a massive central body in a galactic nucleus
and interacting with a thin Keplerian disc. The solution is in a
parametrical form valid for an {\it arbitrary radial distribution of
density and a very broad range of models of the star-disc interaction}.
Temporal evolution can be given in terms of quadrature
provided the star-disc interaction is specified completely
(in terms of function $\Sigma$).
Our approach can be applied to other situations but the form of
eq.~(\ref{2ten}) is linked to the assumption about interactions,
eq.~(\ref{2one}). Also the situation when the orbit intersects
the disc twice-per-revolution requires a specific form of $\Sigma$
to be given and, most likely, it does not allow a complete
analytic solution.

Our solution thus describes the initial phases of the stellar
capture (large eccentricity) and it matches smoothly the
low-eccentricity approximation. Apart from an interesting form of
analytical expressions, our approach is useful as a part of more
elaborate calculations. In an accompanying detailed paper,
additional effects are taken into account (e.g., gravity of the disc)
and  distribution of a large number of stars is investigated
(Vokrouhlick\'y \& Karas 1997, submitted to MNRAS).

We thank the referee for comments concerning temporal evolution of
the orbits and for other suggestions which helped us to improve
our contribution. We acknowledge support from the grants GA\,CR
205/\linebreak[2]97/\linebreak[2]1165 and GA\,CR
202/\linebreak[2]96/\linebreak[2]0206 in the Czech Republic.


\label{lastpage}
\end{document}